\documentclass{ws-procs9x6}

\begin{document}

\title{Interpolating the Free Energy Density Differences of
reweighting methods}

\author{P. R. Crompton}

\address{Institut f\"{u}r Theoreticshe Physik, Universit\"{a}t
Regensburg, D-93040 Regensburg, Germany} 
\address{E-mail: peter.crompton@physik.uni-regensburg.de}


\maketitle

\abstracts{A discussion of the overlap problem of 
reweighting approaches to evaluating critical phenomenon in
fermionic systems is motivated by highlighting 
the divergence of the joint probability density function of a general
ratio. By identifying the bounds for which this integral can
be expressed in closed form, we establish criteria for accurately 
mapping the joint ratio distribution of two disjoint ensembles through 
interpolation. The approach is applied 
to QCD with four staggered flavours to evaluate the critical line
in the $\beta-\mu$ plane.}

\section{Introduction}
The pathology of fermionic reweighting schemes 
can be succinctly expressed in the free energy density 
difference, $\Delta f$, given by the ratio of two partition functions 
for a given finite system of intensive and extensive variables, $\beta$, 
and, $V$. 
Reweighting methods attempt to effect this difference through numerical
simulation by normalising 
observables with the ratio of the Monte Carlo functional
measures for the two separate ensembles. 

\begin{equation}
\frac{\Omega}{\Omega'} = exp\{ \,- \Delta f \,\, \beta  V \, \} \label{f}
\end{equation}

However, as the free energy density difference that can be evaluated 
is necessarily positive the exponent of this ratio can become 
vanishingly small. The relevance of the ensemble that can be numerically
evaluated, $\Omega'$, to the phase space of the ensemble of
interest, $\Omega$, therefore becomes questionable. Simple-mindedly
we could ask how close two ensembles must be for a reweighting 
measurement to be unaffected by this finite difference. 
\section{Probability Density Function of a Ratio}

This question can be expressed in general terms through 
the joint probability density function, $\phi_{\frac{A}{B}}(t)$, 
of the ratio of two normally distributed variables, $x$, and, $y$. 
Both having a given mean, $\mu \,$, and variance, $\sigma$, and in the 
following example below vanishing probability 
densities everywhere but for positive values of $y$ and $x$. 

\begin{equation}
t = \frac{y}{x}
\,\,\,\,\,\,\,\,\,\,\,\,\, y:\phi_{A}( \mu_{A},
\sigma_{A} ) \,\,\,\,\,\,\,\,\,\, x:\phi_{B}( \mu_{B},
\sigma_{B} )
\end{equation}

\begin{eqnarray}
d\phi_{\frac{A}{B}}(t) & = & \int_{x=0}^{\infty} dx \
\phi_{A}(tx).\phi_{B}(x) \int_{y=tx}^{(t+dt)x}dy \\
\phi_{\frac{A}{B}}(t) = 
 \lim_{dt \rightarrow 0} \frac{d\phi_{\frac{A}{B}}(t)}{dt} & = & 
\int_{x=0}^{\infty} \,\, dx \,\, x \,\, \phi_{A}(tx).\phi_{B}(x)
\end{eqnarray}
\begin{eqnarray}
= \frac{\sigma_{A}^{2} \mu_{B}t + \sigma_{B}^{2} \mu_{A}
}{ \sqrt{2 \pi} ( \sigma_{A}^{2} t^{2} + \sigma_{B}^{2})^{3/2}} 
exp \left\{ -\frac{ (\mu_{B} - \mu_{A}t
)^{2} }{ 2 ( \sigma_{B}^{2} + \sigma_{A}^{2}t^{2} )} \right\} \label{2gauss}
\end{eqnarray}

Even for this overly simplistic case the joint probability density 
function of the ratio has a nonzero Cauchy 
component and so all higher moments (including the mean
and variance) are undefined. Insight is gained
with the central limit theorem under Lyapunov conditions
\cite{Kallenberg}, where the distribution of the ratio will 
asymptotically approach normal if the mean of the ratio is several standard
deviations from zero. Similarly, under these conditions the higher moments 
can thus be properly defined for the discrete finite free energy density
 distribution of Eq. (1) which relates to the pathology of reweighting
\cite{noclarice}.

\begin{eqnarray}
\left\langle{ {\mathcal{O}} }\right\rangle_{\beta} 
& = & \frac{\sum_{i}^{N} {\mathcal{O}} e^{-\Delta\beta E_{i}}}
{\sum_{i}^{N} e^{-\Delta\beta E_{i}}} \\
\{ {\mathcal{O}} \} = \frac{1}{N} \sum_{i}^{N}
 {\mathcal{O}}(E_{i}) & e^{-\Delta\beta E_{i}} & 
\,\,\,\,\,\,\,\,\,\,\,\,\,\, \{ {\mathcal{O}},1 \} =
 \frac{1}{N} \sum_{i}^{N}
{\mathcal{O}}(E_{i}) e^{-2\Delta\beta E_{i}} 
\end{eqnarray}

\begin{eqnarray}
\frac{( \delta {\mathcal{O}} )^{2}} {{\left\langle{ {\mathcal{O}}
}\right\rangle_{\beta}}^{2}} & = & \left(  \frac{ \{ {\mathcal{O}} , {\mathcal{O}} \} }{ \{ {\mathcal{O}} \}^{2} } - 1 \right)  \,\, + \,\, \left(
\frac{ \{ 1,  1 \} }{ \{ 1 \}^{2} } - 1 \right)  \,\, - \,\, 2 \left(
\frac{ \{ {\mathcal{O}}, 1 \} }{ \{ {\mathcal{O}} \} \{ 1 \} } - 1
\right) \label{hometime}
\end{eqnarray}

Ignoring the integrated autocorrelation times of the Monte Carlo evaluation 
in the above defintion 
the relative error of a reweighted observable
$\left\langle{ {\mathcal{O}}}\right\rangle_{\beta}$ is clearly
minimised for when the observable ${\mathcal{O}}$ approaches
unity. Conversely, if an observable is normalised so that the mean is
unity with a small standard deviation, we have ensured that
Lyapunov-type conditions are valid (at least for the numerator). 
Since with reweighting a measurement can be redefined 
relative to a different ensemble simply by explicitly
evaluating a free energy density difference in the measurement, the
remaining denominator can be expressed as the product of a series of 
terms each incrementally close to unity. It is then 
straightforward to show with Eq. (\ref{hometime})
 that the relative error of such a product converges with 
an increasing number of increments for the given ${\mathcal{O}}$ 
factored in this manner. By expressing the free energy density
difference of Eq. (1) as the product of series of vanishingly small 
differences under appropriate contraints the finite free energy density 
difference of reweighting can thus be interpolated. 
\section{Sign Problem}

The importance sampling evaluation procedure of a Monte Carlo method is 
essentially undefined for non-positive definite weights. 
This is the case for spin systems such as the Hubbard model where one
effective measure which is used for simulations is the modulus ensemble 
\cite{wiese}.

\begin{equation}
\left\langle {\mathcal{O}} \right\rangle 
= \frac {\int DU \,\, {\mathcal{O}} \,\, 
{\rm{det}M}  e^{-S} } {\int DU \,\, {\rm{det}M} e^{-S} }
\end{equation}

\begin{equation}
{\rm{det}M \,\,\, is \,\,\, real,} \,\,\,\,\,\,\,\,\,\,\,\,
sgn({\rm{det}M}) = \left\{ \begin{array}{c} -1 \\ +1 \end{array} 
\right. \end{equation}

\begin{eqnarray}
\left\langle {\mathcal{O}} \right\rangle_{\|} & = &
\frac {\int DU \,\, {\mathcal{O}} \,\, 
{\| \rm{det}M \|}  e^{-S} } {\int DU \,\, {\| \rm{det}M \|} e^{-S} }\\
\left\langle {\mathcal{O}} \right\rangle & = &
\frac{\left\langle \, {\mathcal{O}}.sgn({\rm{det}M}) \, 
\right\rangle_{\|}}{\left\langle \, sgn({\rm{det}M}) \,
\right\rangle_{\|}} \label{smelly}
\end{eqnarray}

To interpolate $sgn({\rm{det}M})$ 
we would therefore rewrite the ratio of expectations in 
Eq. (\ref{smelly}) as a product
of expectations of increments close to unity. 
Bringing the numerator into a similar product 
form as the denominator if required by a significant probability 
density of the numerator at zero.

\begin{equation}
\frac{\mathcal{O}}{\mathcal{O}_{o}} \sim 1
\,\,\,\,\,\,\,\,\,\,\,\,\,\,\,\,\,\,\,\,\,
{\mathcal{O}_{o}} \, - 1 = N\delta 
\end{equation}
\begin{eqnarray}
\left\langle {\mathcal{O}} \right\rangle & = &
\frac{ \left\langle
\frac{\mathcal{O}}{\mathcal{O}_{o}} \, .\, sgn({\rm{det}M}) \,
\right\rangle_{\|} } { \prod_{n=1}^{N} Re \left\langle
\frac{ 1 + \delta (n-1) }{ 1 + \delta n} \, .\, exp \left\{ \, \frac{i
\pi}{2N} ( 1 + sgn({\rm{det}M})) \, \right\} \,
\right\rangle_{\|} } \label{sign}
\end{eqnarray}

\section{Overlap Problem}

Following the inclusion of the chemical 
potential $\mu$ into the fermionic action, lattice QCD is
similarly unamenable to direct Monte Carlo treatment as
${\rm{det}M(\mu)}$ is complex valued for $\mu \neq 0$. A case in
point is the finite density Glasgow method \cite{phil}. The 
distribution of a set of normalised expansion
coefficients is wanted at a point on the critical
line, $\mu_{1}$, but an 
ensemble can only be generated on the real line for 
$\mu_{o}=0$. We may now, though, express the
relation between the two normalised ensemble-averaged expansion
terms for these two regions ($\mu_{1}, \mu_{o}$) in terms analogous
 to the sign/modulus relation defined for the sign problem in Eq. (\ref{sign}). 

\begin{equation}
\left\langle {\frac{{c_{n}}}{{\rm{det}} M(\mu_{1})}}
 \right\rangle_{\mu_{1}}
 \longleftrightarrow  \left\langle {\frac{{c_{n}}}{{\rm{det}} M(\mu_{o})}}
 \right\rangle_{\mu_{o}} \,\,\,\,\,\,\, 
\left\langle {\mathcal{O}} \right\rangle 
 \longleftrightarrow \left\langle {\mathcal{O}} \right\rangle_{\|} 
\end{equation}

\begin{equation}
\left\langle {\frac{{c_{n}}}{{\rm{det}} M(\mu_{1})}}
 \right\rangle_{\mu_{1}}  
= \frac{\left\langle {\frac{c_{n}}{ {\rm{det}} M(\mu_{1}) } \frac{ {\rm{det}} M(\mu_{1}) }{ {\rm{det}} M(\mu_{o}) }}
\right\rangle_{\mu_{o}} }{ 
 \left\langle {\frac{ {\rm{det}} M(\mu_{1}) }{ {\rm{det}} M(\mu_{o})
}} \right\rangle_{\mu_{o}} } \label{sayswat} 
\end{equation}

As before to accurately map the free energy density difference between 
ensembles a normalising factor is inserted to bring the numerator
close to unity and the remaining free energy
density difference is interpolated in the denominator.

\begin{eqnarray}
\left\langle {\frac{{c_{n}}}{{\rm{det}} M(\mu_{1})}}
 \right\rangle_{\mu_{1}}
& = & 
\frac{ \left\langle {  \frac{c_{n}}{ {\rm{det}} M(\mu_{0})
} \frac{ 1 }{{\rm{det}} M(\mu_{2}) }} \right\rangle_{\mu_{o}}}
{ \left\langle {  \frac{ {\rm{det}} M(\mu_{1})
}{ {\rm{det}} M(\mu_{0})
} \frac{ 1 }{{\rm{det}} M(\mu_{2}) }} \right\rangle_{\mu_{o}}} \\ =  
\frac{ \left\langle {  \frac{c_{n}}{ {\rm{det}} M(\mu_{3})
}} \right\rangle_{\mu_{o}}}
{ \left\langle {  \frac{ {\rm{det}} M(\mu_{1})
}{ {\rm{det}} M(\mu_{3})
} } \right\rangle_{\mu_{o}}} & = & \frac{\left\langle {\frac{c_{n}}{ 
{\rm{det}} M(\mu_{3}) }}
\right\rangle_{\mu_{o}} }{ 
 \prod_{n=1}^{N} \left\langle {\frac{ {\rm{det}} M(n \Delta\mu) }{
{\rm{det}} M( \, [n+1]\Delta\mu \, ) }} \right\rangle_{\mu_{o}} } 
\end{eqnarray}

\begin{equation}
\frac{ c_{n} }{ {\rm{det}} M(\mu_{o})  {\rm{det}} M(\mu_{2})} = 
\frac{ c_{n} }{ {\rm{det}} M(\mu_{3}) } \sim 1 \,\,\,\,\,\,\,\,\,\,\,\,\,\,\,
\frac{{\rm{det}} M(\mu_{1})}{{\rm{det}} M(\mu_{3})}
= {\rm{det}} M(N \Delta \mu)
\end{equation}

For convenience, rather
than determine the normalised distributions at the
critical line from ensembles generated for several $\beta$ values 
 we additionally reweight in $\beta_{1}$. The modified
coefficient being again normalised to unity 
through a choice for $\mu_{3}$. 

\begin{equation}
c_{n} \rightarrow c_{n} \, . \, exp\{ -S_{g}(\beta_{1}) + S_{g}(\beta_{o}) \}
\end{equation}

\begin{figure}[ht]
\centerline{\epsfxsize=4.5in\epsfbox{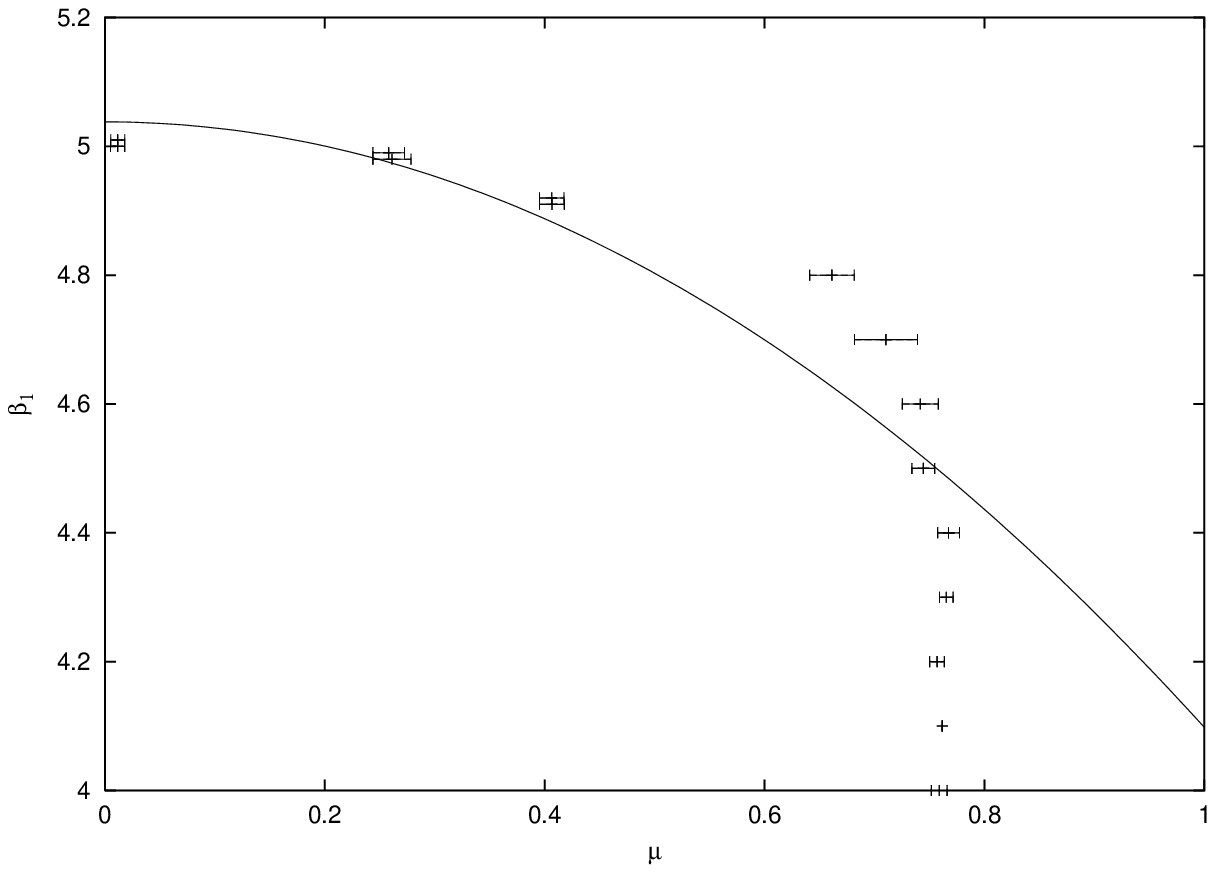}}  
\caption{The critical line in the $\beta_{1}-\mu$ plane  for $n_{f}=4$ QCD, 
determined from the zeros of a fugacity polynomial expansion.
\label{plot}}
\end{figure}

The zeros 
of this polynomial which approach the real axis of
the expansion variable in the thermodynamic limit identify the 
critical line \cite{LeeYang}. A preliminary critical value for $\mu$ is 
plotted as a function of
$\beta_{1}$ in Figure \ref{plot}. It should be
noted that since the polynomial is deflated during rootfinding
there is no actual dependence on the transition value
of $\mu_{1}$, and 
consequentially no need to tune the reweighting parameters 
to effect cancellations through the covariance. 
The above ensemble
consists of 2,000 configurations on a $4^{4}$ volume at 
$\beta_{o}=5.04$ with $m=0.10$, though strictly the quadratic fit 
\cite{Maria} is for a smaller bare mass $(m=0.05)$. 
The congruence is therefore qualitative, 
although the fall-off at $\mu \sim 0.7$ \cite{me} is perhaps more
consistent with expectations for 
the density of nuclear matter at $\beta_{1}=0.0$.

\end{document}